\documentclass[aps,preprint,preprintnumbers,amsmath,amssymb,
  floatfix]{revtex4}

\usepackage{graphicx}  % needed for figures
\usepackage{dcolumn}   % needed for some tables
\usepackage{bm}        % for math
\usepackage{amssymb}   % for math
\usepackage{amsmath}
\usepackage{subfigure}
\usepackage{color}
 \usepackage{slashed}
\usepackage{tikz}
\usepackage{cancel}

\newcommand{\be}{\begin{equation}}
\newcommand{\ee}{\end{equation}}
\newcommand{\bea}{\begin{eqnarray}}
\newcommand{\eea}{\end{eqnarray}}
\newcommand{\bee}{\begin{eqnarray*}}
\newcommand{\eee}{\end{eqnarray*}}

\usepackage{tikz}
\usetikzlibrary{arrows,shapes}
\usetikzlibrary{trees}
\usetikzlibrary{matrix,arrows}              % For commutative diagram
\usetikzlibrary{positioning}                % For "above of=" commands
\usetikzlibrary{calc,through}               % For coordinates
\usetikzlibrary{decorations.pathreplacing}  % For curly braces
\usepackage{pgffor}                         % For repeating patterns
\usetikzlibrary{decorations.pathmorphing}   % For Feynman Diagrams
\usetikzlibrary{decorations.markings}
\tikzset{
    vector/.style={decorate, decoration={snake}, draw},
    provector/.style={decorate, decoration={snake,amplitude=2.5pt}, draw},
    antivector/.style={decorate, decoration={snake,amplitude=-2.5pt}, draw},
    fermion/.style={draw=black, postaction={decorate},
        decoration={markings,mark=at position .55 with {\arrow[draw=black]{>}}}},
        fermion1/.style={draw=black, postaction={decorate}},
    fermionbar/.style={draw=black, postaction={decorate},
        decoration={markings,mark=at position .55 with {\arrow[draw=black]{<}}}},
    fermionnoarrow/.style={draw=black},
    gluon/.style={decorate, draw=black,
        decoration={coil,amplitude=4pt, segment length=5pt}},
    scalar/.style={dashed,draw=black, postaction={decorate},
        decoration={markings,mark=at position .55 with {\arrow[draw=black]{>}}}},
    scalar1/.style={dashed,draw=black, postaction={decorate}},
    scalarbar/.style={dashed,draw=black, postaction={decorate},
        decoration={markings,mark=at position .55 with {\arrow[draw=black]{<}}}},
    scalarnoarrow/.style={dashed,draw=black},
    electron/.style={draw=black, postaction={decorate},
        decoration={markings,mark=at position .55 with {\arrow[draw=black]{>}}}},
    bigvector/.style={decorate, decoration={snake,amplitude=4pt}, draw},
}

\begin{document}

\title{Alternative search strategies for a BSM resonance fitting ATLAS diboson excess}

\author{Biplob Bhattacherjee\footnote{biplob@cts.iisc.ernet.in}, Pritibhajan Byakti\footnote{pritibhajan@cts.iisc.ernet.in}, Charanjit K. Khosa\footnote{khosacharanjit@cts.iisc.ernet.in}, \\ 
Jayita Lahiri\footnote{jayita@cts.iisc.ernet.in} and  Gaurav Mendiratta\footnote{gaurav@cts.iisc.ernet.in} }

\affiliation{Centre for High Energy Physics, Indian Institute of
Science, Bangalore- 560012, India}

\date{\today}

\begin{abstract}
We study an s-channel resonance $R$ as a viable candidate to fit the diboson excess reported by ATLAS.
We compute the contribution of the $\sim 2$ TeV resonance $R$ to semileptonic and leptonic final states  at 13 
TeV LHC. To explain the absence of an excess in semileptonic channel, we explore the possibility where the particle $R$ 
decays to additional light scalars $X,X$ or $X,Y$. Modified analysis strategy has been proposed to study three particle 
final state of the resonance decay and to identify decay channels of $X$. Associated production of $R$ with gauge bosons has been studied in detail
to identify the production mechanism of $R$. We construct comprehensive categories for vector and scalar BSM particles which 
may play the role of particles $R$, $X$, $Y$ and find alternate channels to fix the new couplings and search for these 
particles.

\end{abstract}

%\pacs{}

 \maketitle
 \newpage
\section{Introduction}
Resonant searches in the s-channel mediated $2\to2$ process are special as they can provide a smoking gun signal for beyond standard model (BSM) at the LHC. ATLAS collaboration has
recently reported a $2.6\sigma - 3.4\sigma$ excess in their searches for BSM resonances in diboson channel decaying into two fat jets final states
  from 20.3 fb$^{-1}$ of data at 8 TeV LHC\cite{atlasdiboson1}. Any hints in the electro-weak (EW) sector in TeV energy are exciting as almost all the major ultraviolet (UV) completions of Standard Model (SM) predict TeV scale particles. The observed invariant mass distribution of the cross section in diboson channel shows a bump in 1.8-2 TeV region. This gives further support
       to the weak evidence of an excess of (1$\sigma $-2$\sigma$) which were already seen in the same
        channels by the CMS experiments in a similar invariant mass region\cite{cmshadronic}.

Since the excess arises at ~2 TeV scale, the final state gauge bosons($W/Z$) are highly boosted and collimated. In the hadronic decay of a boosted gauge boson, the light quark jets produced are also highly collimated and form one fat jet ($J$). While looking for the two fat jet final
  states whose jet mass peaks around the $W/Z$ mass, ATLAS
   found the largest discrepancy near 2 TeV corresponding to 3.4 $\sigma$, 2.6 $\sigma$ and 2.9 $\sigma$ significance in $WZ$, $WW$ and $ZZ$ channels
     respectively. Considering entire mass range 1.3-3.0 TeV in each of the search channel, the global
      significance of the discrepancy in the WZ channel is 2.5$ \sigma$.

Any BSM particle which decays to two standard model gauge bosons
$W/Z$ will also have semileptonic and leptonic final states
because of the leptonic decay modes of $W$ and $Z$. Therefore corresponding to the dijet channel, the
semileptonic and leptonic channels should also see some excess. However, no such excess has been observed in
  the searches by CMS or ATLAS\cite{cmssemileptonic,cmslllnu,Aad:2015ufa,Aad:2014ry,Aad:2014pl}.
    In the combined analysis of  hadronic, semileptonic and leptonic final states, ATLAS finds that the largest deviation from the background expectation is 2.5$ \sigma $ and corresponds to a 2 TeV invariant mass\cite{atlasdiboson2}. This significance is smaller than the 3.4 $ \sigma$ significance observed in
      the JJ channel as the semi-leptonic and leptonic channels are consistent with the background-only hypothesis. 
  We should comment here that the leptonic branching fraction of $W$ and $Z$ are very small, for example in the fully leptonic decay channel, $ZZ$ $\rightarrow$ $4 l$, is not very competitive in the diboson resonance search at 8 TeV.

In other related searches, ATLAS and CMS collaborations have presented their results and put upper limits on the $Zh/Wh$ production cross
section \cite{Khachatryan:2014pqr,Khachatryan:2015zxd,Aad:2014lkh}. CMS collaboration observes a
2.2 $\sigma$ excess over
 the standard model at $m_{Wh}$ $\approx$ 1.8 TeV in the $l \nu b \bar b$ final state\cite{Khachatryan:2015zxd}.
 Dijet resonance searches also show a (2.1$\sigma$) discrepancy in the invariant mass region of
  $\approx$2 TeV\cite{Aad:2014aqa,Khachatryan:2015sja}. Individually these searches have a low statistical
   significance and may get ruled out with early LHC run II\cite{Goncalves:2015yua}. At the same time,
   the combination of all of these excesses and the appealing signature of TeV scale BSM physics
    encourage us to look for more ways look for this TeV scale BSM physics in the early LHC run-II.

Naturally, any hint of a discrepancy between SM predictions and
observations creates a renewed excitement and a flurry of
suggestions for possible explanations and suggestions for further
explorations. The possible models which may explain partly or
fully, the above experimental results include additional new gauge
bosons such as $W'$,$Z'$\cite{WZprime},
composite Higgs models\cite{composite}, heavy Higgs boson(s)\cite{Chaostealth,Chen2tevres,heavyhiggs}, string originated
\cite{Stringyorigin} and $R$-parity violating SUSY model\cite{Allanach:2015blv}, walking technicolor \cite{Fukano:2015hga} and other spin one resonances \cite{spin1}. In
effective field theory approach for bosonic production
is used to study diboson channel\cite{dibosoneft}.

In this study we explore comprehensively, the production, decay and coupling measurement of a BSM s-channel scalar or vector bosons which can fit the diboson excess.
    In section \ref{sec1} we will review the all experimental results from ATLAS and CMS
     involving the diboson search. Then, for analysis at
13 TeV LHC, we follow the same strategy as adopted by ATLAS to estimate the number of events in production
of $WW,WZ$ and $ZZ$ and their decays to jets, semi-leptonic and leptonic channels. To study BSM models,
we start with discussing status of two particle final states coming from the decay of a heavy resonance. Then we check
the viability of three particle final state as a possible mimic for the diboson excess. In section \ref{sec2}, explore
the production mechanism of the heavy resonance to look for probes which distinguish between quark-quark or gluon-gluon
initiated processes. We propose alternate channels and cuts to distinguish these two cases in the associated production
process. Next we comment on how to independently measure the couplings of the BSM resonance with gauge bosons and quarks. In section \ref{sec3}, we discuss
the possibility that, as in the case of 8 TeV, the semi-leptonic and leptonic channels do not show an excess even in the 13 TeV run but
in the hadronic decay mode the excess survives. An additional light BSM particle which can mimic the $W/Z$ boson signatures at LHC would
be favoured in this scenario. We explore such a model and find general signatures of its decay modes to isolate the BSM physics. In section \ref{sec4}, we discuss
and categorize simplified models which can accommodate a 2 TeV resonance and additional $W/Z$-like BSM particles with
mass $\sim 100$ GeV. In the last section, we discuss some general experimental signatures common to many of the proposed BSM simplified models. In case, this resonance
is verified and the LHC run-II confirms its existence with greater statistical significance, following the strategy described in our paper, its couplings and decay modes
can be identified in the early LHC run.

\section{Analysis \label{sec1}}
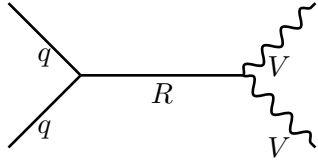
\begin{figure}
 \centering
\begin{tikzpicture}[line width=1.0 pt, scale=1.2]

\draw[fermion1] (0,0.0)--(0.8,-0.8); \draw[fermion1]
(0,-1.6)--(0.8,-0.8);

% \draw[dashed] (3.6,-0.8) -- (4.2,0.0);
     \node at (0.4,-.6) {$q$};
    \node at (0.4,-1.4) {$q$};
     \node at (1.7,-1.0) {$R$};

     \node at (3.0,-1.6) {$V$};
     \node at (3.0,-0.7) {$V$};

    \draw[fermion1] (0.8,-0.8) -- (2.6,-0.8);
   \draw[vector] (2.6,-0.8) -- (3.4,-0.0);
   \draw[vector] (2.6,-0.8) -- (3.4,-1.6);
\end{tikzpicture}
\vspace{.5cm} \caption{Feynman diagram for resonance decaying to
two vector bosons.} \vspace{1.0cm} \label{feyndiag}
\end{figure}

In this section we will review ATLAS diboson analysis and CMS
results for this diboson excess. ATLAS experiment at LHC has
reported diboson excess around 2 TeV in the hadronic decay of
gauge bosons\cite{atlasdiboson1} only. They do not see any excess
in leptonic channels in their updated
analysis\cite{atlasdiboson2}.
 Since gauge bosons coming from the heavy resonance will be very boosted, so their
  decay products will be highly collimated. Two quarks from hadronic
   decay of $W/Z$ will form few highly collimated jets which then look like a single jet of bigger radius and these
    objects are called fat jets. ATLAS has scanned the
invariant mass range of $1.3-3$ TeV to look for any BSM
resonances decaying to the diboson channel. As the final states
are fat jets, in their analysis, jets are constructed using C/A
method with radius 1.2. After jet formation, a grooming
algorithm, a variant of mass drop technique, is applied to find
pair of pair of quarks forming fat jet and to reduce pileup. Mass
drop technique is used to examine the sequence of pairwise
combinations used to reconstruct the jet to find two subjets
corresponding to the $W$ or $Z$ boson decay identified in
simulated signal events. Feynman diagram for this process is given in Fig. \ref{feyndiag}.
 For semileptonic and leptonic decay
channels, in order to maximize the sensitivity to resonances with
different masses, three different optimized set of selection
criteria are used according to the $p_T$ of the leptonically
decaying $W(p_T^{l \nu})$, $Z(p_T^{ll})$ and hadronically decaying
$W/Z$ ($p_T^{jj}$ or $p_T^J$ where $p_T^J$ means the $p_T$ of the
fat jet). These three regions are low $p_T$ resolved, high $p_T$
resolved  and merged region. As the resonance is 2 TeV, its decay products will be highly boosted and $p_T$ of boosted objects will be greater than 400 GeV and high $p_T$ merged
region is most suitable for this scenario. CMS has not observed any excess and their upper bound on
cross-section is given in Table \ref{cmscrossr}. We will discuss
ATLAS analysis in detail, results or bounds for each channel one
by one.
\subsection{Resonance decaying to $WW$, $ZZ$ or $WZ$}
\subsubsection{JJ channel} ATLAS has observed some excess in this channel from 2 TeV resonance decay.
 This channel is sensitive to $WW$, $ZZ$ and $WZ$ processes. They look for two fat jets with 69.4
$<m_J^{W}<$ 95.4 GeV and 79.8 $<m_J^{Z}<$ 105.8 GeV,  $p_T^J>$
540 GeV, $|\eta|<$ 2 and
$\frac{p_T^{J_1}-p_T^{J_2}}{p_T^{J_1}+p_T^{J_2}}$ $<$ 0.15
($p_T^{J_1}$ and $p_T^{J_2}$ denote $p_T$ of first and second jet
respectively). Only those events are selected when number of
charged tracks inside the jets are less than 30 and 
$\slashed E_T$ is less than 350 GeV. Jets having subjets less than
3 are considered. Fat jet pair invariant mass is demanded greater
than 1050 GeV. QCD dijet is the largest background for $JJ$
channel and is removed by tagging the vector bosons with
\begin{enumerate} \item $\sqrt{y}>0.5$ for sub-jets (before
re-clustering)
 \item Number of charged tracks $<30$ for
ungroomed jets to remove higher multiplicity gluons \item
$m_{J}\in\left(82.4\pm13||92.8\pm13\right)$GeV to tag $W/Z$ bosons
\item $|y_{1}-y_{2}|<1.2$ between 2 leading jets to remove the
large t-channel gluon mediated background \end{enumerate} In
addition to the above cuts, the ATLAS analysis removed events with
a prompt electron ($E_{T}>20$ GeV, $|\eta|<1.37$ or
$1.52<|\eta|<2.47$) or a muon ($p_{T}>20$ GeV, $|\eta|<2.5$) to
avoid overlap with leptonic decay searches in diboson channel and
also removed events with $\slashed E_T >$ 350 GeV to remove overlap
with searches with $Z\to\nu\overline{\nu}$. Limit on diboson
cross-section for all the processes corresponding to 2 TeV
resonance from ATLAS  are given in Table \ref{atlascrossr}.
\begin{table}[htb]
 $$
 \begin{array}{|c|c|c|c|}
 \hline{\mbox{Upper bound on cross-section}}&{\mbox{WW}}& ZZ & WZ \\
 \hline

JJ & 16  & 18 & 16 \\
l\nu J & 3.5 & - & -  \\
llJ & - & 8.5 & - \\
lll \nu & - & - & 0.3  \\

\hline
  \end{array}
 $$
\caption{CMS 8 TeV diboson bounds on cross-section in fb ($R \rightarrow VV $) for different channels
\cite{cmslllnu,cmssemileptonic,cmshadronic}.}
 \label{cmscrossr} \end{table}
\begin{table}[htb]
 $$
 \begin{array}{|c|c|c|c|}
 \hline{\mbox{Upper bound on cross-section}}&{\mbox{WW}}& ZZ & WZ \\
 \hline
JJ & 30   & 30  & 30  \\
l\nu J & 5  & - & 9 \\
llJ & - & 8  & 20  \\
lll \nu & - & - & 21  \\
\mbox{Combined (hadron + leptonic)} & 13   & 13 & 15 \\
\hline
  \end{array}
 $$
\caption{ATLAS 8 TeV diboson bounds on cross-section ($R \rightarrow
VV$) (in fb) for different channels corresponding to $m_R$ $~$ 2
TeV. \cite{atlasdiboson1,atlasdiboson2}.}
 \label{atlascrossr} \end{table}
\subsubsection{l$\nu$J channel}
ATLAS has updated analysis of diboson excess also includes leptonic
decays of gauge bosons \cite{atlasdiboson2}. One can get this
final state either from $WZ$ or $WW$ decay.  Events which have only
one isolated (high $p_T$) lepton fall under this category.
Following cuts have been used :
\begin{enumerate}
\item  Jet invariant mass range : $65<m_W<105$ GeV and
$70<m_Z<110$ GeV \item Lepton $p_T$ $>$ 25 GeV   \item  Jet
$p_T^J$ $>$ 400 GeV \item $ \slashed E_T$ $>$ 30 GeV   \item
$p_T(l \nu)$ $>$ 400 GeV (vector sum of $p_T$ vector of $\slashed
E_T$ and lepton)
\end{enumerate} Main backgrounds in this
channel are $W$ $+$ jet, top quark pair production and
non-resonant diboson production. $\Delta \phi$ cut : $\Delta \phi$
$(\slashed E_T,J)$ $>$ 1 is applied to reject multi-jet background.
Top quark production background is reduced by rejecting the events
with b-tagged jet having $\Delta R$ $>$ 0.8 (with fat jet).
\subsubsection{llJ channel}
This channel is sensitive to $ZZ$ and $WZ$ final state only. All
the cuts for this channel are same as previously discussed
channels except the condition of same flavour opposite sign
dilepton pair. Leptonic pair invariant mass ($m_{ll}$) is
required in 65-115 GeV range. Main background processes are $Z$
$+$ jets, top quark pair and non- resonant vector boson pair
production.
\subsubsection{$lll \nu$ channel} Here three isolated leptons each with
$p_T$ $>$ 25 GeV and  $ \slashed E_T>$ 25 GeV are demanded. Out of
these two leptons should be of same flavour opposite sign.
Invariant mass of that pair of leptons should be with in 70-110
GeV ($m_Z \pm $ 20 GeV) range since they can be produced only from
$Z$ decay. This final state one can get only from $WZ$ decay.
Dominant backgrounds for this channel are SM- $WZ$ and $ZZ$
production. Data driven method is used to estimate these
background processes.

ATLAS has observed an excess in the $JJ$ channel of gauge boson
decay. If the resonance is a true signal, beyond statistical
fluctuations, the 13 TeV LHC will be able to see a sharp peak in
the $JJ$ channel because the cross-section will increase significantly 
at 13 TeV. The $JJ$ channel is sensitive to
$WW$, $WZ$ and $ZZ$ equally because of same hadronic branching of
$W$, $Z$. Therefore from this channel we can't distinguish whether
the resonance is dominantly decaying to $WW$ or $WZ$ or $ZZ$.
Hence to comment about its charge we have to be certain about its
decay channels. Moreover along with the $JJ$ channel the other
decay modes of $W$ and $Z$ i.e. the semileptonic and fully
leptonic channels should also observe some clear excess. If the 13
TeV LHC see some excess events in semileptonic or fully leptonic
decay channels. We will be able to distinguish which decay mode of
the resonance is preferred. For example, the $lll \nu$ channel is
only sensitive to $WZ$; $llll$ is sensitive to $ZZ$ only; $l \nu
J$, is sensitive to both $WW$ and $WZ$. But significant events in
$l \nu J $ channel, and no event in $lll \nu$ will ensure that the
final state is $WW$. If they observe no signal in any of the
channels, they will be able to rule out the possibility that the
resonance decays to diboson final state.

\begin{table}[htb]
 $$
 \begin{array}{|c|c|c|c|c|c|}

\hline &\rm { Cuts }& \multicolumn{3}{ |c| }{\mbox{No of }}&\mbox{ No. of } \\
&\rm {  }& \multicolumn{3}{ |c| }{\mbox{signal}}& \mbox{background} \\
&\rm {  }& \multicolumn{3}{ |c| }{\mbox{events}}& \mbox{events} \\
 &\rm {  }& \multicolumn{3}{ |c| }{\rm {(5 fb ^{-1} ) }}& \rm {( 5 fb ^{-1}) } \\
\hline{\mbox{Channel}} & &{\mbox{WW}}& ZZ & WZ & \\
 \hline

l\nu J & p_T^l > 25 \,\,\mbox{GeV}, \,\, p_T^J > 800 \,\, \mbox{GeV}, \,\, p_T^{l \nu} > 800 \,\, \mbox{GeV}  & 65  &  & 30 & 20 \\
\hline llJ & 1.8 \,\, \mbox{TeV} < m_{llJ} < 2.2 \,\, \mbox{TeV},\,\, p_T^J > 400 \,\,\mbox{GeV},\,\, p^T_{ll} > 400 \,\, \mbox{GeV} &  & 30  &  10 & 2 \\
  \hline lll \nu & p_T^{l_1} > 100\,\, \mbox{GeV},\,\, p_T^{l_2} > 100 \,\,\mbox{GeV}, \,\, p_T^{l_3} > 100\,\, \mbox{GeV} &  & & 4 & 4\\
  \hline ll ll & p_T^{l_1} > 60 \,\, \mbox{GeV},\,\,  p_T^{l_2} > 60 \,\, \mbox{GeV},\,\, p_T^{l_3} > 60 \,\,\mbox{GeV}, \,\, p_T^{l_4} > 60 \mbox{GeV} &  & 1 & & 2 \\  \hline
  \end{array}
 $$
\caption{Expected number of events for different leptonic channels at 13
TeV. }
 \label{eventcount13} \end{table}

To check this, we have done a detailed analysis following the same
strategy as ATLAS to calculate the relevant number of events in
each channel which will be observed in the 13 TeV LHC. We have
also estimated the major backgrounds for all these channels. We
used CalcHEP\cite{CalcHEP} for parton level event generation,
Pythia6.4.28\cite{pythia} for showering and fastjet-3.1.3\cite{fastjet} for
jet formation.  Only track isolation is applied to count number of isolated
leptons.  If the $p_T$ sum in cone of radius 0.2 is less than 15
$\%$ of lepton $p_T$ then that lepton is declared as an isolated
lepton. For $ l \nu J$ channel, we have considered $W$ $+$ jets,
$Z$ $+$ jets and SM $VV$ production as main background. In case of
$ll J$, we have considered $Z$ $+$ $J$ and SM $VV$ production as
backgrounds. In case of $llll$, SM $VV $ have been
considered as dominant background.  The results have been
presented in Table \ref{eventcount13}. We have assumed $R$ $\rightarrow$ $VV$ 100 fb cross-section for $WW$ and also for $ZZ$ and $WZ$ at 13 TeV. In principle  these three cross-sections can be different from each other but one can rescale the number of events accordingly. The used optimized cuts to reduce
background and to increase signal significance which are also listed in
the Table \ref{eventcount13}.

Expected number of events for this process from $WW,$ $ZZ$ and
$WZ$ decay modes are given in  Table \ref{eventcount13} for 13 TeV LHC.  We can see that $WW$ channel can be easily 
differentiated from the other modes with only 5 fb$^{-1}$ of data, because in the $WW$ case, there will be an excess only in the $l \nu J$ channel. To see the events in the fully leptonic channel, we need higher luminosity $ \sim$ 100 fb$^{-1}$. In case we
 find some excess in $ll J$ channel but not in $l \nu J$ channel that will definitely correspond to $ZZ$ final state. If one finds excess in both $l \nu J$ and $ll J$ channel then it will be the $WZ$ final state.

 \begin{figure}
 \centering
\begin{tikzpicture}[line width=1.0 pt, scale=1.0]

\draw[fermion1] (0,0.0)--(0.8,-0.8); \draw[fermion1]
(0,-1.6)--(0.8,-0.8);

% \draw[dashed] (3.6,-0.8) -- (4.2,0.0);
     \node at (0.4,-.6) {$q$};
    \node at (0.4,-1.4) {$q$};
     \node at (1.7,-1.0) {$R$};

     \node at (3.0,-1.4) {$V_1$};
     \node at (3.0,-0.2) {$Y$};

\node at (4.0,0.2) {$V_2$}; \node at (4.0,-0.6) {$X$};
    \draw[scalar1] (0.8,-0.8) -- (2.6,-0.8);
   \draw[scalar1] (2.6,-0.8) -- (3.2,-0.4);
   \draw[vector] (3.2,-0.4) -- (3.8,-0.0);
   \draw[scalar1] (3.2,-0.4) -- (3.8,-0.8);
   \draw[vector] (2.6,-0.8) -- (3.8,-1.6);
\end{tikzpicture}
\vspace{.5cm} \caption{Feynman diagram for resonance decaying to
two vector bosons and some new particle.} \vspace{1.0cm}
\label{2to3process}
\end{figure}
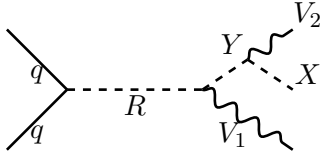

\begin{table}
 $$
 \begin{array}{|c|c|c|c|c|}
\hline M_Y & M_X & p_T^A < .15  & p_T^A < .25 &  p_T^A < .5 \\
{\small{ \mbox{(in GeV)}}} & {\small{ \mbox{(in GeV)}}} & && \\
 \hline
 150.0   &     10.0  &  0.28 &
0.28   &   0.28  \\

200.0 &  50.0  &  0.13 & 0.13 &
0.14  \\

300.0 &  50.0 &   0.16 & 0.17 &
0.17  \\

 300.0 &  100.0  &  .063 &
.063  &    .064  \\

400.0  &  200.0  & .0003 & .0003 & .0003\\ \hline
  \end{array}
 $$
\caption{Fraction of events passing different $p_T^A$ asymmetry
cuts
$\left(\frac{p_T^{V_1}-p_T^{V_2}}{p_T^{V_1}+p_T^{V_2}}\right)$,
1.8 $< m_{V_1 V_2} < $ 2.2 TeV  and $X$ decays invisibly.}
 \label{xdecayinginvisi} \end{table}

\begin{table}
 $$
 \begin{array}{|c|c|c|}
\hline{ {M_Y}} &  {M_X} & \mbox{{\small{Fraction of the events}}} \\
{\small{ \mbox{(in GeV)}}} & {\small{ \mbox{(in GeV)}}} & {\small{ \mbox{with}\,\, \Delta R < 1.2 }} \\
 \hline
 150.0   &      10.0   &
 0.97 \\
 200.0    &    50.0    &   0.97
 \\
 300.0    &    50.0  &    0.86
 \\
 300.0   &     100.0   &    0.92
 \\
 400.0    &    200.0  &     0.87
 \\  \hline
  \end{array}
 $$
\caption{Fraction of events having $X$ decay products within fat
jet radius, for different values of $Y$ and $X$ mass.}
 \label{xdecayingvis} \end{table}

\begin{figure}
\centering
\includegraphics[scale=0.35]{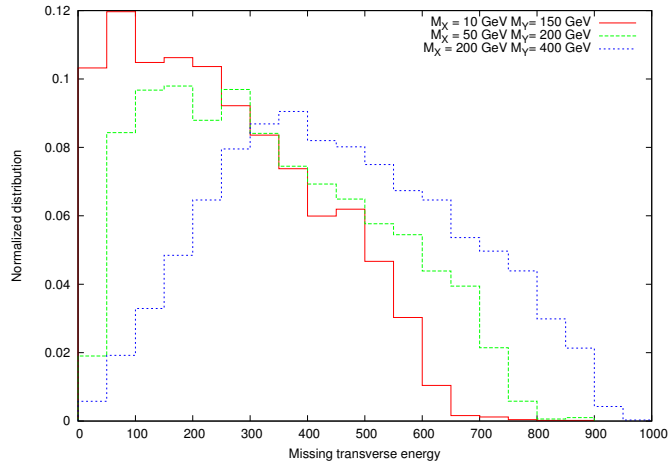}
\caption{$\slashed E_T$ distribution when $X$ decays invisibly for three different sets of $X$ and $Y$ masses as shown in the inset.}\label{miset2to3}
\end{figure}

So far we have considered 2 $\rightarrow$ 2 topology. Recently it
 has been argued in \cite{saavedra} that resonance (which we denote by $R$) can also
 decay to three particles in the final state as shown in Fig. \ref{2to3process}.
The schematics of this type of process is 
$pp$ $\rightarrow$ $V_1 V_2 X$. We investigated the viability of this scenario. To this end, we generate parton level events for $R$ decaying to 3 body final
state. For this process we have three new particles : $X$, $Y$ and $R$, including the resonance.
We want to explore the
possibility of this final state mimicking the diboson final state.
The particle $X$ can decay into visible or to invisible final states. It can
 not decay to leptons because that final state is not very difficult to identify.
In case $X$ decays invisibly, there will be a $p_T$ asymmetry
between the visible final states. Hence we calculate the fraction
of events respecting different jet $p_T$ asymmetry cuts for
different combinations of $Y$ and $X$ particle masses. We have
taken a few benchmark points for $X$ and $Y$ masses. We have
explicitly put the invariant mass cut of 1.8 TeV$<m_{V_{1}V_{2}}<$2.2 TeV
on the visible final state particles. We can see from the Table
\ref{xdecayinginvisi} that it is very difficult to satisfy $p_T$
asymmetry condition unless $X$ is very light. For $M_X$=10 GeV, only 27
$\%$ of the events pass the $p_T$ asymmetry cut. But even if this is the
case, one can  confirm this decay chain by looking at the two fat jets $+$ missing $\slashed{E_T}$
signal (see Fig.
\ref{miset2to3}).

\begin{figure}
\centering
\includegraphics[scale=0.35]{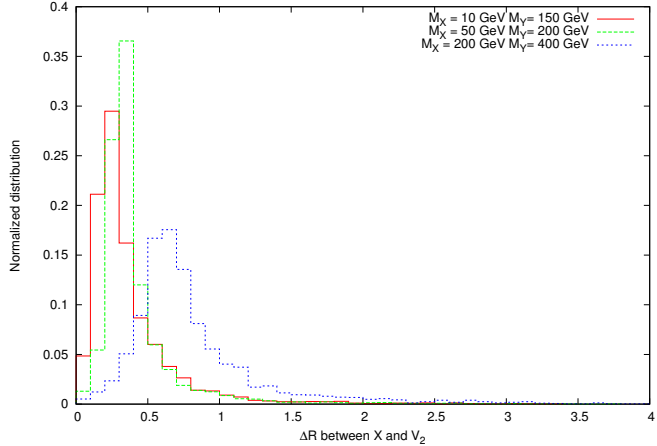}
\caption{$\Delta R$  distribution between $X$ and $V_2$ for three different sets of $X$ and $Y$ masses as shown in the inset.}\label{del2to3}
\end{figure}

Second possibility is when $X$ decays to hadrons. Again
we look for fraction of events when $X$ lies within ( $\Delta R$
$<$ 1.2 ) fat jet radius (see Fig. \ref{del2to3}). We can see from the Table
\ref{xdecayingvis} that most of events will pass this condition.
Even with high $X$ mass 87 $\%$ of the events will have $X$ and
$V_2$ ($W$ or $Z$) lying within fat jet radius. In that case, the invariant
mass of the two fat jets will also peak around 2 TeV. But still there
is a way to rule out this possibility. One of the fat jet
invariant mass will peak at $M_Y$ instead of gauge boson mass.
Low $X$ mass option might pass this criteria. It will be
within 1.2 radius and $p_T$ asymmetry cut will also not be able to
rule out this possibility. But in that case, the jet substructure inside the fat jet 
will be different. To identify this
scenario one has to slightly modify the mass drop technique to
look for two parents instead of one and to identify the jet
substructure fully when some of the components are not very
massive. There can be another possibility that the $X$ lies
outside the 1.2 radius and decays into visible particles. But that
probability is only about 10$\%$. So we do not consider that
possibility here. So if $X$ decays invisibly then low mass $X$ can can pass $p_T$ 
asymmetry cut used by ATLAS and in that case one should look for two fat jets + $\slashed E_T$ signal.
For visible $X$ decay, ungroomed jet mass $M_Y$. One will find more substructure, and more number of tracks inside the fat jet. 
In that case one should follow different strategy to look for jet substructure.

\section{Associated Resonance Production \label{sec2}}
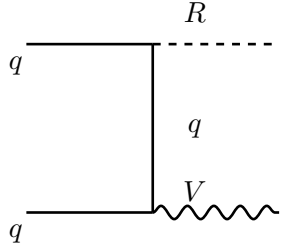
\begin{figure}
 \centering
\begin{tikzpicture}[line width=1.0 pt, scale=1.4]

\draw[fermion1] (4.5,0.0)--(5.7,0.0); \draw[scalar1]
(5.7,0.0)--(6.9,0.0); \draw[fermion1] (4.5,-1.6)--(5.7,-1.6);
     \node at (4.4,-.2) {$q$};
    \node at (4.4,-1.8) {$q$};
     \node at (6.1,0.3) {$R$};
     \node at (6.1,-1.4) {$V$};
     \node at (6.1,-0.8) {$q$};
  \draw[fermion1] (5.7,0.0) -- (5.7,-1.6);
   \draw[vector] (5.7,-1.6) -- (6.9,-1.6);
\end{tikzpicture}
\vspace{.5cm} \caption{Feynman diagram for quark initiated  $RV$
production processes.} \vspace{1.0cm} \label{feyndiagasso}
\end{figure}

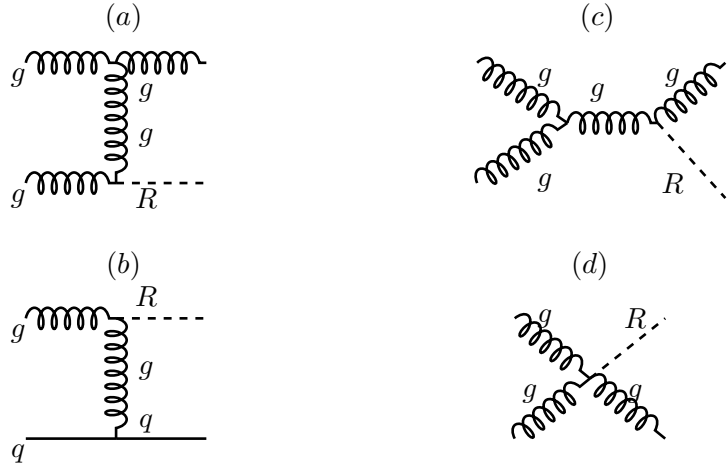
\begin{figure}
 \centering
\begin{tikzpicture}[line width=1.0 pt, scale=1.0]
\node at (5.8,1.0) {$(a)$}; \draw[gluon] (4.5,0.4)--(5.7,0.4);
\draw[gluon] (5.7,0.4)--(6.9,0.4); \draw[gluon]
(4.5,-1.2)--(5.7,-1.2);

     \node at (4.4,0.2) {$g$};
    \node at (4.4,-1.4) {$g$};
     \node at (6.1,0.0) {$g$};
     \node at (6.1,-0.6) {$g$};
     \node at (6.1,-1.4) {$R$};
   \draw[gluon] (5.7,0.4) -- (5.7,-1.2);
   \draw[scalar1] (5.7,-1.2) -- (6.9,-1.2);

%second diagram
\node at (12.1,1.0) {$(c)$};
 \draw[gluon] (10.5,0.4)--(11.7,-0.4);
\draw[gluon] (10.5,-1.2)--(11.7,-0.4);

\draw[gluon] (11.7,-0.4)--(12.9,-0.4);

\draw[gluon] (12.9,-0.4)--(13.8,0.4);

\draw[scalar1] (12.9,-0.4)--(13.8,-1.4);
  \node at (11.4,0.2) {$g$};
    \node at (11.4,-1.2) {$g$};
     \node at (12.1,0.0) {$g$};
     \node at (13.1,0.2) {$g$};
     \node at (13.1,-1.2) {$R$};

%third diagram
\node at (5.8,-2.3) {$(b)$}; \draw[gluon] (4.5,-3.0)--(5.7,-3.0);
\draw[scalar1] (5.7,-3.0)--(6.9,-3.0);
 \draw[fermion1](4.5,-4.6)--(5.7,-4.6);

     \node at (4.4,-3.2) {$g$};
    \node at (4.4,-4.8) {$q$};
     \node at (6.1,-2.7) {$R$};
     \node at (6.1,-3.7) {$g$};
     \node at (6.1,-4.4) {$q$};
   \draw[gluon] (5.7,-3.0) -- (5.7,-4.6);
   \draw[fermion1] (5.7,-4.6) -- (6.9,-4.6);

%fourth diagram
\node at (12.0,-2.3) {$(d)$}; \draw[gluon]
(11.0,-3.0)--(12.0,-3.8);

\draw[gluon] (11.0,-4.6)--(12.0,-3.8);

\draw[scalar1] (12.0,-3.8)--(13.0,-3.0);

\draw[gluon] (12.0,-3.8)--(13.0,-4.6);

\node at (11.4,-3.0) {$g$};
    \node at (11.2,-4.0) {$g$};
     \node at (12.6,-3.0) {$R$};
     \node at (12.6,-4.0) {$g$};

\end{tikzpicture}
\vspace{.5cm} \caption{Feynman diagrams for associated dijet
production.} \vspace{1.0cm} \label{gluoass}
\end{figure}

Having discussed the 13 TeV projections of the various leptonic decay modes
of the newly observed resonance, we expect that the LHC run II can
easily see the resonance through these channels also, and can
identify its decay modes precisely. Then one will have clear
understanding of decay modes of the 2 TeV resonance. Our next task
is to address the issue of production mechanism of this state.
Question arises whether the production of the resonance $R$ is a quark
initiated or a gluon initiated process. Quarks have tree level gauge 
couplings with electroweak gauge bosons but these couplings do not exist for gluons. 
Therefore for quark initiated
process we can get gauge bosons $W^{\pm}$, $Z$, $\gamma$ and $g$
(see Fig. \ref{feyndiagasso} for Feynman diagrams) emission from
the initial quark legs. Associated production of $R$ with $W,Z, \gamma$ will ensure that the 
initial state is quark. Gluon has tree level couplings with both quark and gluon.
Hence a gluon jet can come from both quark or gluon leg in the initial state
initial state \footnote{In \cite{Liew:2015osa} associated jet production channel has been considered where the resonance decays invisibly.}(see Fig. \ref{gluoass} (a)), therefore associated
gluon production with $R$ can not distinguish the initial states.
But if gluon coupling with $R$ exists 
we will get associated quark jet with $R$ from the process depicted in Fig. \ref{gluoass} (b)
also. Using quark/ gluon tagging it may be possible to distinguish
these two diagrams. As $W^\pm$, $Z$, $\gamma$ can only come from
quark initiated process, any signature of the associated
production of $R$ with $W^\pm$, $Z$ and $\gamma$ will ensure that
$R$ is produced through quark coupling. Schematic of associated
production is as follows : \be
 qq \rightarrow RV \rightarrow WW+ W/Z/\gamma \rightarrow JJ + l \nu/ll/ \gamma \quad
 ; \quad
   qq \rightarrow R \rightarrow WW \ee Typical signals one should look for $RW$ and $RZ$ production at 13
TeV LHC are $JJ l \nu$ and $JJ l^+ l^-$.

\begin{figure}
\centering
\includegraphics[scale=0.6]{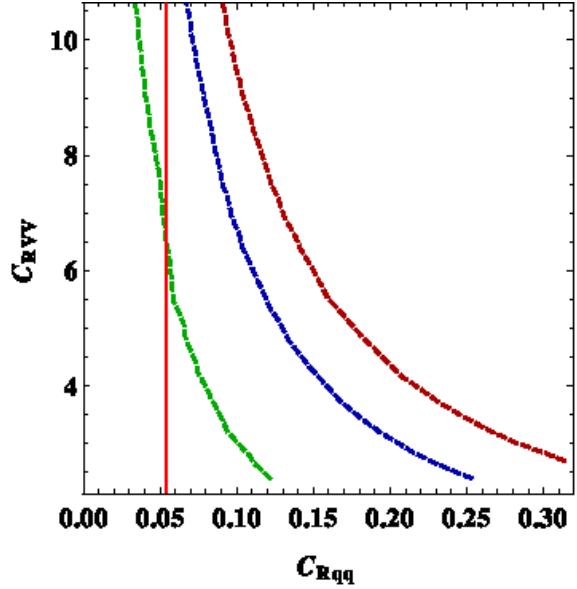}
\caption{Contours of constant $\sigma (pp \rightarrow R
\rightarrow WW)$(fb) in $C_{Rqq}$ $C_{RVV}$ plane. The maroon, blue and
 green dotted curves denote 10 fb, 5 fb and 1 fb cross-section contours respectively. The red
solid line is the projected dijet bound at 14 TeV. }\label{abcont}
\end{figure}

From this discussion it is evident that associated production of
$RV$ is an important channel to explore at the 13 TeV LHC. Hence
we will estimate the cross-section for this process. For that one
has to specify the model. There have been many analyses based upon
spin 0, spin 1 and spin 2 resonance, to explain diboson excess. We
consider in our work, a spin 0 resonance, produced through quark
initiated process. Having spin information we can find out the
viable model parameter space in context of the reported excess.
Models with scalar $R$ involve only two couplings i.e. coupling of
$R$ with vector bosons ($C_{RVV}$) and R coupling with quarks
($C_{Rqq}$) for $pp \rightarrow R \rightarrow VV$ process. The
coupling $C_{Rqq}$ also determines the dijet cross-section. From
the dijet resonance searches by ATLAS and CMS at 8 TeV we find
that $\sigma (pp \rightarrow R \rightarrow jj) < 100$ fb for a 2
TeV resonance \cite{Aad:2014aqa,Khachatryan:2015sja}.  Our aim is to look for parameter space consistent with both
diboson and dijet constraints. We find that to satisfy dijet
cross-section limit $C_{Rqq} < 0.33$. As mentioned the process $pp
\rightarrow R \rightarrow VV$ involves both $C_{Rqq}$ and
$C_{RVV}$ coupling and its cross-section is proportional to
product of these two couplings.
The ATLAS upper limit on the diboson cross section is $\approx 10$
fb \cite{atlasdiboson1} at 8 TeV (We use average of cross-section limits from CMS and
ATLAS). 
In Fig. \ref{abcont} we show contours of constant $\sigma (pp
\rightarrow R \rightarrow VV)$ in the $C_{Rqq}$ and $C_{RVV}$
plane. The whole parameter space shown in Fig. \ref{abcont} is compatible
with the 8 TeV dijet
constraints\cite{Aad:2014aqa,Khachatryan:2015sja}.  The red dotted
line denotes the contour which gives the diboson $ \sigma (pp
\rightarrow R \rightarrow VV )$ $\approx$ 10 fb which is the upper
limit on the diboson cross section from ATLAS. So parameter space
below this curve is allowed by diboson search. We have plotted the
projected limit for 14 TeV LHC for the dijet cross section \cite{Yu:2013wta}. The red solid line in the Fig. \ref{abcont} is the
projected dijet limit for 14 TeV. Only the region to the left of
this line will be allowed by the dijet search at 14 TeV. Assuming $Z'_B$ model, 14 TeV limits on $Z'$-quark coupling has been calculated in \cite{Yu:2013wta}. Using this limit, we have calculated 14 TeV projected limit on dijet cross-section which we used to extract the 14 TeV limit on the $C_{Rqq}$ coupling in our model, corresponding to 300 fb$^{-1}$ luminosity. We can
see from the figure that 14 TeV LHC can rule out most of the
currently allowed parameter space.

 We have chosen
 few benchmark points $(C_{Rqq},C_{RVV})$ from the allowed
 parameter space.
In Table \ref{djandadj13est} we quote the 13 TeV associated $RV$,
dijet and associated dijet cross-section for the benchmark points.

\begin{figure}
 \subfigure[]{
\includegraphics[scale=0.3,angle=-90]{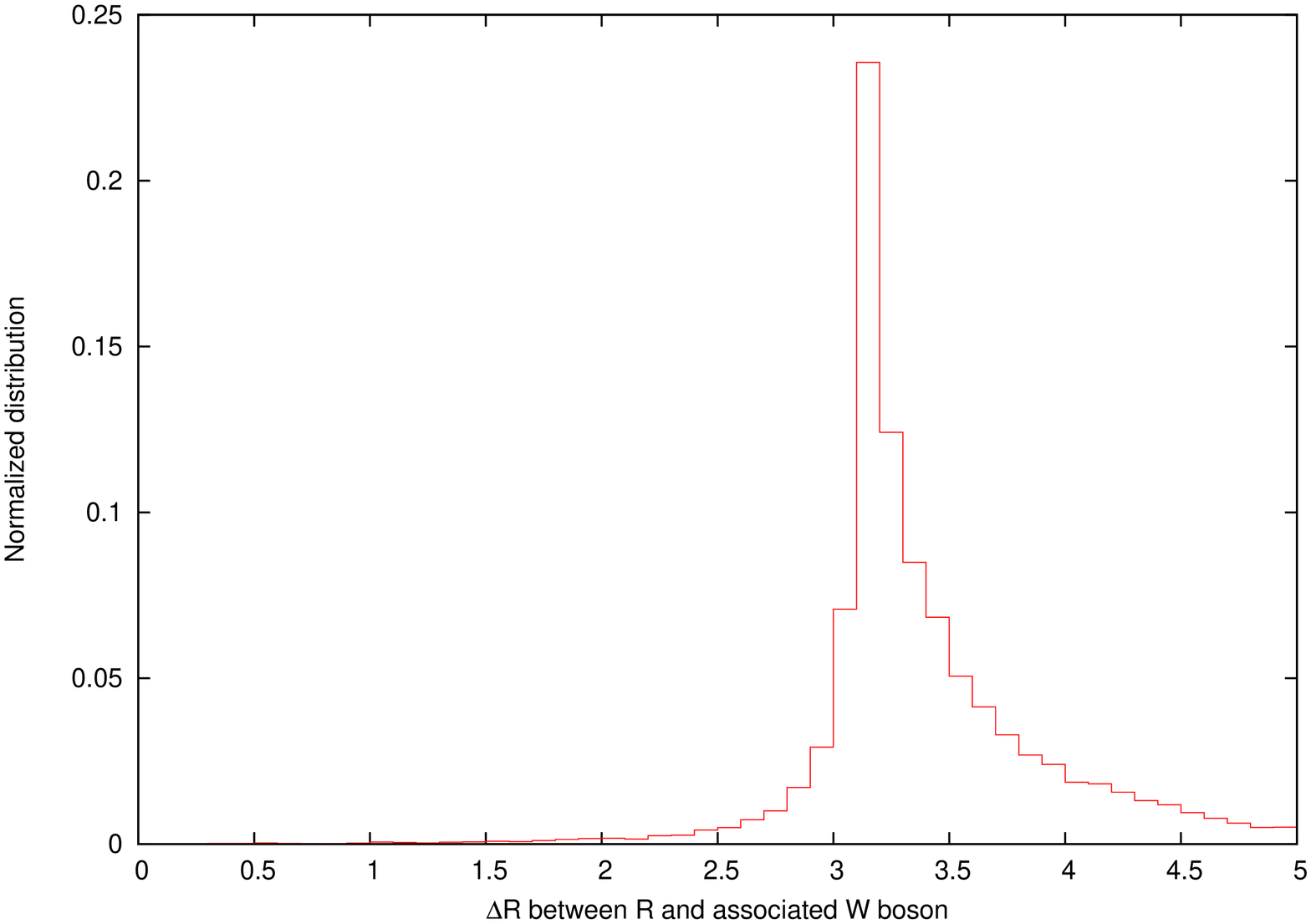}}
 \subfigure[]{
\includegraphics[scale=0.3, angle = -90]{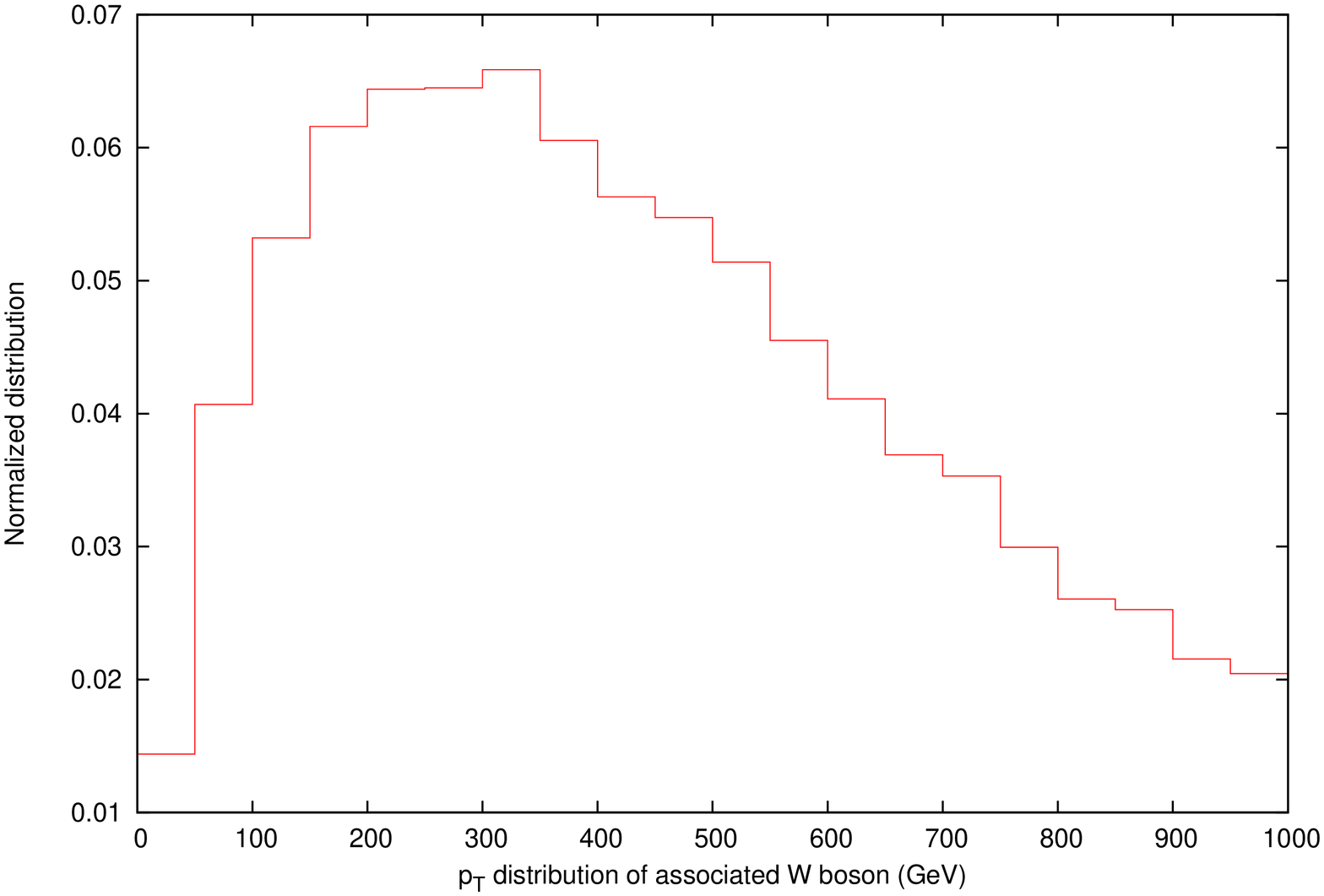}}
\caption{(a) $\Delta R$  distribution between $R$ and
$W$, (b) $p_T$ distribution of $W$  produced in association with
$R$.}\label{ptlepassorw}
\end{figure}

Now our objective is to present a projection of the associated
production of $R$ with gauge bosons for 13 TeV LHC. Using first
benchmark($C_{Rqq}=0.3$, $C_{RVV}=2.67$) point we have generated 20,000 parton level events for $ pp
\rightarrow RW$ and $RZ$ processes,  using CalcHEP. We pass these
events to Pythia 6.4.28 for hadronization and used fastjet for jet
formation. Main backgrounds for these processes are $W/Z$ $+$
jets. To ensure that the lepton(s) coming from
$W$/$Z$ are isolated and well separated from the decay products of
the high $p_T$ gauge bosons, we plot $\Delta R$ distribution
between $W$ and $R$ in Fig. \ref{ptlepassorw} (a). $W$ boson $p_T$ distribution for associated $RW$
is given in Fig. \ref{ptlepassorw} (b). One can see from these two
figures that the lepton will be indeed isolated and it will carry enough $p_T$ such that this can be
a clear signal to identify at LHC.  We estimate the number of events for $JJ l \nu$ and $JJ ll$
processes for 13 TeV LHC run. All our cuts are same as $JJ$
channel of ATLAS diboson analysis except for the $JJ$ invariant
mass cut which we consider in the range 1800-2200 GeV in this
case. Diboson signal strength will be maximum in the narrow region
around 2 TeV. For associated $RW$ production in the $JJ l \nu$
final state we have used the following cuts :

\begin{table}
 $$
 \begin{array}{|c|c|c|c|}

 \hline{{C_{Rqq},C_{RVV}}}& (0.3,2.67) & (0.09,10.13) & (0.11,7.99) \\
 \hline

\sigma (RW ) & 75.6 \,\, \mbox{fb} & 6.8 \,\,  \mbox{fb} & 10.26 \,\,  \mbox{fb} \\
\sigma (RZ) & 32.4 \,\,  \mbox{fb} &  3.02 \,\,  \mbox{fb} & 4.42 \,\, \mbox{fb} \\
\sigma (R \gamma) & 4.24 \,\, \mbox{fb}\,\,  ( p_T > 20\,\,  \mbox{GeV} ) & 0.72 \,\, \mbox{fb}\,\,
 (p_T > 20 \,\,\mbox{GeV}) & 1.14 \,\, \mbox{fb}\,\, (p_T > 20 \,\, \mbox{GeV}) \\

\sigma \mbox{(dijet )}& 470.0 \,\, \mbox{fb} & 1.44 \,\,  \mbox{fb} & 4.8 \,\, \mbox{fb}\\
\sigma \mbox{(associated dijet)} & 57.6 \,\, \mbox{fb} & 0.16 \,\, \mbox{fb} & 0.6 \,\, \mbox{fb} \\
\hline
  \end{array}
 $$
\caption{13 TeV cross-section estimate for different processes for three benchmark points.}
 \label{djandadj13est}
\end{table}

\begin{enumerate}
\item $ 30  < p_T^l < 350$ GeV  \item  $ 30  < \slashed {E_T} <
350$ GeV
\end{enumerate}
For associated $RZ$ decaying to $JJll$ final state only the lepton
$p_T$ cut is used.  Expected number of events
for $JJ l \nu$ and $JJll$ are given in Table
\ref{eventcountassorw} and \ref{eventcountassorz} respectively.

\begin{table}[htb]
 $$
 \begin{array}{|c|c|c|}
 \hline{{\sqrt{s}}}&{\rm{ 8 TeV}} &{\rm  13 TeV} \\
 \hline
 {\rm {Total\,\, No.\,\, of \,\, Generated\,\, Events }} & 20000 & 20000  \\ \hline
\sigma(pp \rightarrow RW)  & 5 \,\, \mbox{fb}  & 75.6  \,\, \mbox{fb}  \\ \hline
 JJ l \nu & 277 & 161 \\
\hline
  \end{array}
 $$
\caption{Expected number of events for $JJl \nu$ from $RW$
associated production using the first benchmark points for couplings $C_{Rqq}$=0.3 and $C_{RVV}$=2.67.}
 \label{eventcountassorw} \end{table}

\begin{table}[htb]
 $$
 \begin{array}{|c|c|c|}
 \hline{{\sqrt{s}}}&{\rm{ 8 TeV}} &{\rm  13 TeV} \\
 \hline
 {\rm {Total\,\, No.\,\, of \,\, Generated\,\, Events }} & 20000 & 20000  \\ \hline
\sigma (pp \rightarrow RZ)  & 2.2 \,\, \mbox{fb}  & 32.0 \,\, \mbox{fb}  \\ \hline
 JJ ll  & 101 & 69 \\
\hline \end{array}
 $$
\caption{Expected number of events for $JJ ll$ from $RZ$
associated production using using the first benchmark points for couplings $C_{Rqq}$=0.3 and $C_{RVV}$=2.67.}
 \label{eventcountassorz} \end{table}

If the resonance is produced through quark initiated
process then 13 TeV LHC will be able to confirm that looking at
associated production of $R$ with gauge bosons in the 2 fat jets 
$+$1 lepton $+\slashed{E_T}$ in case of associated $RW$ and a pair of leptons with 2 fat jets in case of associated $Z$ production.
$RW$ is the best channel to look for, since $W$ branching to
lepton is much higher than $Z$ branching ratio to the leptons
which is clear from the Table \ref{eventcountassorw} and
\ref{eventcountassorz}. The backgrounds for these processes are found to be negligible. 
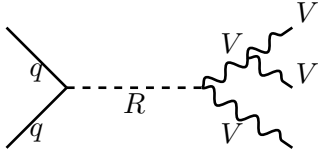
\begin{figure}
 \centering
\begin{tikzpicture}[line width=1.0 pt, scale=1.0]

\draw[fermion1] (0,0.0)--(0.8,-0.8); \draw[fermion1]
(0,-1.6)--(0.8,-0.8);

% \draw[dashed] (3.6,-0.8) -- (4.2,0.0);
     \node at (0.4,-.6) {$q$};
    \node at (0.4,-1.4) {$q$};
     \node at (1.7,-1.0) {$R$};

     \node at (3.0,-1.4) {$V$};
     \node at (3.0,-0.2) {$V$};

\node at (4.0,0.2) {$V$}; \node at (4.0,-0.6) {$V$};
    \draw[scalar1] (0.8,-0.8) -- (2.6,-0.8);
   \draw[vector] (2.6,-0.8) -- (3.8,-0.0);
   \draw[vector] (3.2,-0.4) -- (3.8,-0.8);
   \draw[vector] (2.6,-0.8) -- (3.8,-1.6);
%    \draw[fermion] (4.0,-1.8) --(4.8,-1.8);
%     \draw[vector] (4.2,-1.7) -- (4.7,-1.1);
\end{tikzpicture}
\vspace{.5cm} \caption{Feynman diagram for resonance decaying to
three vector bosons.} \vspace{1.0cm} \label{assorvbck}
\end{figure}

Although we have argued that looking at associated production of $R$ with gauge bosons is 
is the best strategy to identify the production mechanism of $R$, we should also mention that
a 2 $\rightarrow$ 3 process where R decays to $WWZ$ (s-channel diagram - see Fig.
\ref{assorvbck}) will give us the same signature as the associated
production $RZ$ (where $R (\rightarrow WW) Z $) or  $RW$ ($(R
\rightarrow WZ) W $). This process can come from quark quark as well as gluon gluon final states. 
Hence looking at only associated production of $R$ with gauge bosons is not enough to comment about the production
mechanism of $R$.
The typical cross section for this type of
processes is 0.5 fb and 4 fb approximately at 8 TeV and 13 TeV
respectively for the first benchmark point($C_{qqR}$=0.3 and $C_{RVV}$=2.67. 
We can see from Table \ref{djandadj13est} we can see that the cross section 
of this process is much smaller than the typical $RW$ or $RZ$ associated production cross section.
There are one more efficient way to distinguish between these two cases.
In addition to
looking for leptonic signals with two fat jets, one should also
count the number of positive and negative charged leptons in this
process. We will always find some asymmetry in the lepton number
count if it is a quark initiated process. The reason behind this
is that the cross-section of $u \bar{d} \rightarrow R W^+$ and $d
\bar{u} \rightarrow R W^-$ is different in proton-proton collider.
Absence of total lepton charge asymmetry will reflect the
possibility that we are looking at three gauge boson final state,
produced through $R$ decay, through s-channel production of $R$ (
see Fig. \ref{assorvbck}). Hence looking at the associated production of $R$ with $W$ as well as 
counting the number of positive and negatively charged leptons one can be definite about the production process of $R$.

The s-channel production of $RW$ or $RZ$ which also has the same final state as the associated production
is not important, because its cross section is of $\cal$ $10^{-5}$ fb.

As one of our objective to extract the couplings $C_{qqR}$ and $C_{RVV}$, we have 
looked at two other processes which involve such couplings.
Diboson production gives an upper bound on the product of $C_{qqR}$ and $C_{RVV}$ as discussed earlier. 
We have given dijet and associated dijet production cross section for a few benchmark points. These
two processes depend only on the $C_{qqR}$ coupling and if we can extract this coupling from their cross section we 
will be able to comment on $C_{RVV}$ coupling. 
Feynman diagrams for associated dijet production are
given in Fig. \ref{gluoass}. We should comment here that the dijet production process has large QCD background. 
Hence coupling measurement from this channel has a lot of uncertainty. 
Associated dijet production is contaminated with less background. Hence this channel is better to extract couplings.

\section{Resonance decaying to two BSM particles \label{sec3}}
\begin{figure}
 \centering
\begin{tikzpicture}[line width=1.0 pt, scale=1.0]

\draw[fermion1] (9,0.0)--(9.8,-0.8); \draw[fermion1]
(9,-1.6)--(9.8,-0.8);

% \draw[dashed] (3.6,-0.8) -- (4.2,0.0);
     \node at (9.4,-.6) {$q$};
    \node at (9.4,-1.4) {$q$};
     \node at (10.7,-1.0) {$R$};

     \node at (12.0,-1.6) {$X$};
     \node at (12.0,-0.7) {$X$};

    \draw[fermion1] (9.8,-0.8) -- (11.6,-0.8);
   \draw[vector] (11.6,-0.8) -- (12.4,0.0);
   \draw[vector] (11.6,-0.8) -- (12.4,-1.6);

   \draw[fermion1] (12.4,0.0) -- (13.0,0.75);
   \draw[fermion1] (12.4,0.0) -- (13.2,0.55);
   \draw[fermion1] (12.4,-1.6) -- (13.2,-1.85);
   \draw[fermion1] (12.4,-1.6) -- (13.1,-2.05);
    \node at (12.6,0.6) {$j$};
     \node at (13.0,0.0) {$j$};
\node at (13.0,-1.6) {$j$};
     \node at (12.8,-2.2) {$j$};

\end{tikzpicture}
\vspace{.5cm} \caption{Feynman diagram for resonance decaying to
pair of particle $X$. } \vspace{1.0cm} \label{feyndiagrxx}
\end{figure}
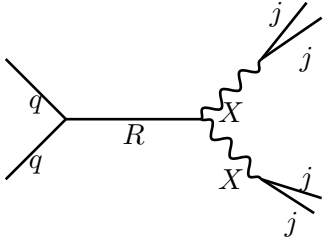

In section \ref{sec2}, we have analyzed the number of events in
all the leptonic final states coming from the decay of $WZ$, $WW$
and $ZZ$. The semi-leptonic and fully leptonic final
states being extremely clean we show the expected signal in
these channels at the 13 TeV LHC in Table ~\ref{eventcount13}. Observation of these events will confirm that a resonance actually decays into a pair
of gauge bosons. If we do not see any excess in the semi-leptonic
and leptonic decay channels of gauge boson at 13 TeV and the hadronic signature of BSM appears again, it would rule out a model which
decays to $W/Z$ bosons alone. In such a case, an additional particle $X$ (see Fig. \ref{feyndiagrxx}) with a mass in the 70-100 GeV range which decays
to light quarks/gluons will be required to explain this excess. The possible decays of particle $X$ are discussed below,
\begin{enumerate}
\item It can decay to light quarks or gluons. \item $b \bar{b}$
final state. In which case, with b-tagging of the fat jets
one will be able to identify this final state. We should comment
here that the b-tagging efficiency will be less for high $p_T$
jets ($ < 50 \% $)\cite{btagging}. Still, it is possible to count number of
b-subjets inside fat jets.

\item $X$ can also decay into pair of $\tau$ leptons. This decay
mode can not be dominant decay channel of $X$ otherwise we will
see some excess in the leptonic channels of diboson searches through tau decay. We can
observe ditau tagged (boosted) fat jets and identify this final
state also.
\end{enumerate}

\begin{figure}
\centering
\includegraphics[scale=0.40]{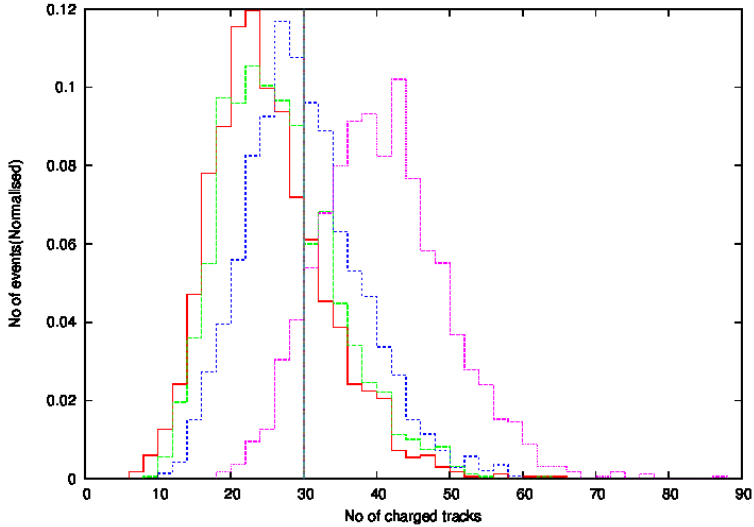}
\caption{Normalized charge multiplicity distribution for highest
$p_T$ jet formed by different ($u \bar{u}$- red, $c \bar{c}$-
green, $b \bar{b}$- blue and $g g$- violet) decay modes of
particle $X$.}\label{chargedis}
\end{figure}
We have studied in detail different scenarios, where $X$ can decay into gluons or quarks.
Using Pythia6.4.28, tune (AUET-2B) and pdf-CTEQ6L1, we have simulated the listed decay
channels of the particle $X$. In Fig. \ref{chargedis}, we show
number of charge tracks  in the fatjet with $p_T$ $>$ 540 GeV
coming from $u \bar{u}$, $c \bar{c}$, $ b\bar{b}$ and gluon -gluon
final states for comparison. If it decays to gluons then one
finds more number of charge tracks in a fat jet as compared to
quark jets\cite{quarkgluon}. As the ATLAS experiment puts a cut that number of charge tracks $<30$ inside a fat jet, the
gluon final state as should not be the dominant decay mode of $X$ to avoid reduction in the cross section via the
branching fraction. Additionally, with the experimental cuts, signal selection efficiency in case of gluon jets is
very small (20 $\%$) and we need high cross-section to observe this final state. It is very difficult
to distinguish $u \bar u $, $c \bar c$ and $b \bar b$ final states from charge track distribution, unless the b-jets are tagged. 
As mentioned earlier, in the high $p_T$ regions, b-tagging efficiency
is lower. Once we have enough data, we can see b-jet sub-jets in the fat jets. When one $X$
decays to quarks and the other to gluons, this scenario can be
identified within low luminosity because quark-quark final state
has more signal selection efficiency, as shown in the Fig.
\ref{chargedis}.

The particle $X$ can be tracked via its decay modes  discussed above at early LHC run-II. The
gluon fraction in the final state can be checked by changing
(increasing) the number of charge track cut on the fat jet. For dijet final state the QCD background also becomes large
when this cut is relaxed but associated production scenario, discussed
in the previous section because of less background it may form a viable channel for the final state analysis.

\section{Anatomy of models explaining diboson excess \label{sec4}}
 In order to fit to the bump observed in the data, s-channel 
models are favored. In the following section, 
we will explore a number of viable s-channel BSM models and discuss them in the 
context of diboson excess and associated processes at LHC.

\begin{figure}
 \centering
\begin{tikzpicture}[line width=1.0 pt, scale=1.0]
%first diagram
dot/.style={circle,fill=black,minimum size=4pt,inner sep=0pt,
        outer sep=-1pt},
\draw[fermion1] (0,0.0)--(0.8,-0.8); \draw[fermion1]
(0,-1.6)--(0.8,-0.8);
     \node at (0.8,-0.1) {$q/g$};
    \node at (0.8,-1.3) {$q/g$};
     \node at (1.7,-1.0) {$R$};
%     \draw[dot] at (1.7,-1.0);
    
     \node at (1.7,-2.4) {$T_1$};
     \draw[fill=black] (0.8,-0.8) circle [radius=0.06];
  \draw[fill=black] (2.6,-0.8) circle [radius=0.06];
    \draw[fermion1] (0.8,-0.8) -- (2.6,-0.8);
   \draw[fermion1] (2.6,-0.8) -- (3.4,-0.0);
   \draw[fermion1] (2.6,-0.8) -- (3.4,-1.6);

%second diagram
\draw[fermion1] (4.0,0.0)--(4.8,-0.8); \draw[fermion1]
(4.0,-1.6)--(4.8,-0.8);
 \draw[fill=black] (4.8,-0.8) circle [radius=0.06];
  \draw[fill=black] (6.6,-0.8) circle [radius=0.06];
     \node at (4.8,-0.1) {$q/g$};
    \node at (4.8,-1.3) {$q/g$};
     \node at (5.7,-1.0) {$R$};

     \node at (5.7,-2.4) {$T_2$};
    \draw[fermion1] (4.8,-0.8) -- (6.6,-0.8);
   \draw[fermion1] (6.6,-0.8) -- (7.0,-0.4);
\draw[fermion1] (7.0,-0.4) -- (7.4,-0.0); \draw[fermion1]
(7.0,-0.4) -- (7.4,-0.8); \draw[fill=black] (7.0,-0.4) circle
[radius=0.06];
   \draw[fermion1] (6.6,-0.8) -- (7.4,-1.6);

%third diagram

\draw[fermion1] (8.0,0.0)--(8.8,-0.8); \draw[fermion1]
(8.0,-1.6)--(8.8,-0.8);
 \draw[fill=black] (8.8,-0.8) circle [radius=0.06];
  \draw[fill=black] (10.6,-0.8) circle [radius=0.06];
     \node at (8.8,-.1) {$q/g$};
    \node at (8.8,-1.3) {$q/g$};
     \node at (9.7,-1.0) {$R$};
  
     \node at (9.7,-2.4) {$T_3$};
    \draw[fermion1] (8.8,-0.8) -- (10.6,-0.8);
   \draw[fermion1] (10.6,-0.8) -- (11.4,-0.0);
   \draw[fermion1] (10.6,-0.8) -- (11.4,-1.6);
   \draw[fermion1] (10.6,-0.8) -- (11.4,-0.8);

\end{tikzpicture}
\vspace{.5cm} \caption{Topologies of s-channel BSM resonance.}
\label{fig:topologies}
\end{figure}
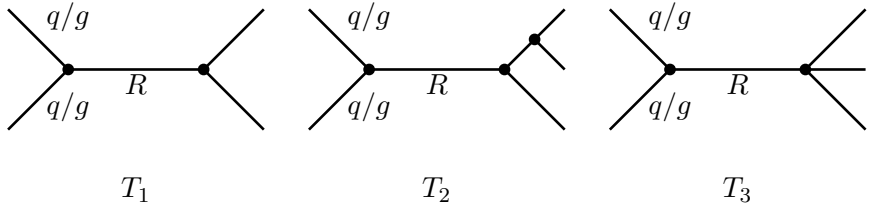

While building s-channel models, we limit to a 2 body or 3 body final states.  
This is because, a larger number of particles in the final state would suppress 
the cross section due to phase space factors and also due to the kinematic 
cuts, 
for example, the requirement of jet $p_T$ asymmetry within 
$15\%$\cite{atlasdiboson1}. Hence, three 
topologies exist which can mimic the excess in the diboson data as shown in 
Fig. 
\ref{fig:topologies}.

In all of the considered models, a 2 TeV particle ($R$) is produced in either  
a 
quark initiated or a gluon initiated process. The mediating particle $R$ can be 
a scalar, vector or a spin 2 boson\footnote{We are not considering this 
possibility in our work.}. Next, we discuss the production of the 
particle $R$ with a gluon or quark initial state.

\subsection{Production}
$R$ can interact with quarks via tree level couplings whereas gluons may  
interact with $R$ via a 5 or 6 dimensional effective operator. Below, we 
discuss 
both of these possibilities in detail.

{\bf Gluon initiated process:} Here the particle $R$ is produced from 
gluon fusion initial state. The Kronecker product $8\times 8$ of the color 
representations of  
gluon initial state dictates that the particle $R$ can belong to a 1, $8$, 
$10$, $\overline{10}$ or $27$ dimensional representation of $SU(3)$ 
\cite{Slansky:1981yr}. Firstly, we  consider the possibility that $R$ is a 
vector boson. The SM gluon clearly will not satisfy the role of $R$ since it is 
massless. The case where the BSM gauge boson is singlet under $SU(3)_C$, the 
lowest dimensional operator for its interaction with the gluon is,
$G_{\mu\nu}G^{\mu\nu} F'_{\alpha\beta}F'^{\alpha\beta}$, where  $  
F'_{\alpha\beta}$ is the field strength of R. Such a particle cannot be 
produced 
singly from gluon initial state and hence cannot belong to any of the 
topologies 
considered.

Next, we consider the case where $R$ is a scalar field.  From the point of view 
of production of a scalar $R$, any of the above color representations are 
allowed. As we will note later, the required final state will rule out all the 
colored representations of a scalar particle $R$. This leaves,  for a 
color-singlet scalar, two choices of effective interactions with the gluon,
\begin{enumerate}
 \item $G_{\mu\nu} G^{\mu\nu} R,$
 \item $G_{\mu\nu}G^{\mu\nu} R^\dagger R$.
\end{enumerate}
In the first case, $R$ is a singlet under SM gauge group.  In the second 
case $R$ may have non-trivial representations under $SU(2)\times U(1)$. 
However, 
in this case, $R$ has to get a non zero vacuum expectation value (VEV) to 
obtain 
a $g g R$ vertex. In addition, tree-level corrections to the $\rho$-parameter 
constraint $R$ to belong to one of the  1, 2 or 7 dimensional 
representation of $SU(2)\times U(1)$\cite{Bhattacharyya:2009gw}.

{\bf Quark initiated:} The case where $R$ is a vector gauge boson, the SM gauge 
group is extended  with 
additional symmetries. In principle, a UV completion of SM can be complicated, 
though, at the TeV scale, BSM gauge particles interacting with quarks have 
similar signatures as $W'$ or $Z'$ models. 
These possibilities have been amply explored in the literature and we  limit to 
the alternative where $R$ is a scalar. For a scalar $R$, possibilities for its couplings with the quarks are listed in 
Tab. \ref{Tab:qRcouplings}. Here, we list up to dimension six operators along 
with the corresponding SM gauge representation of R denoted as 
(SU(3),SU(2),U(1)). The hypercharge listed in the Table \ref{Tab:qRcouplings} is
defined in a convention where Higgs boson has hypercharge $\frac12$. 

%\begin{eqnarray}
\begin{table}
\begin{center}
\begin{tabular}{|c|c|}
\hline
Operator & Representation of $R$\\
\hline
\multicolumn{2}{|c|}{Renormalizable interactions}\\ 
\hline
$Q_L^T C Q_L R$ & $\left(\bar 6,3,\frac13 \right), \left(\bar 6,1,\frac13 
\right), \left(3,3,\frac13 \right), \left(3,1,\frac13 \right)$\\
\hline
$u_R^T C u_R R $ & $\left( 6,\bar 3,-\frac43 \right), \left( 6,1,-\frac43 
\right), \left(\bar 3,3,-\frac43 \right), \left(\bar 3,1,-\frac43 \right)$\\
\hline
$d_R^T C d_R R $ & $\left( 6,\bar 3,\frac23 \right), \left( 6,1,\frac23 
\right), \left(\bar 3,3,\frac23 \right), \left(\bar 3,1,\frac23 \right)$\\ 
\hline
$\bar Q_L u_R R$ & $\left(8,2,-\frac12 \right), \left(1,2,-\frac12 
\right)$\\
\hline
$\bar Q_L d_R R$ & $\left(8,2,\frac12 \right), \left(1,2,\frac12 
\right)$\\ 
\hline
\multicolumn{2}{|c|}{Higher dimensional}\\
\hline
 $\left( \bar Q u_R \tilde H \right) R, \, \left( \bar Q d_R H \right) 
R$ & $\left(8,1,0\right), \, \left(1,1,0\right)$\\
\hline
$\left( \bar Q u_R \tilde H \right) R^\dagger R, \, \left( \bar Q d_R H \right) 
R^\dagger R$  & $\left(1,1,0 \right), 
%\left(1,2,\pm\frac12 \right),
\left(1,7,\pm 2 \right)$\\ 
\hline
 $\left( \bar Q u_R  H \right) R$ & 
$\left(8,3,-1\right), \left(8,1,-1\right), \left(1,3,-1\right) 
\left(1,1,-1\right)$\\
\hline
 $\left( \bar Q d_R \tilde H \right) R$ & $\left(8,3,1\right), 
\left(8,1,1\right), \left(1,3,1\right) 
\left(1,1,1\right)$\\
\hline
$\left( \bar Q u_R \tilde H \right) R^\dagger R, \, \left( \bar Q d_R H \right) 
R^\dagger R$  & $\left(1,1,0 \right), 
\left(1,7,\pm 2 \right)$\\ 
\hline
$\bar{q} \cancel{D} q R$ & $\left(1,1,0\right)$\\
\hline
$\bar{q} \cancel{D} q R^\dagger R$ & $\left(1,1,0 \right), 
%\left(1,2,\pm\frac12 \right),
\left(1,7,\pm 2 \right)$\\ 
\hline
\end{tabular} 
\end{center}
\caption{Quark couplings with a scalar particle $R$}\label{Tab:qRcouplings}
\end{table}

This completes our discussion of the production of the particle 
$R$.

\subsection{Explicit models for $2\to 2$ topology}
The final state particles that $R$ decays into may be SM ($W,Z$) or BSM particles 
of similar mass as $W/Z$ bosons. We call the class of models where the final state 
particles are SM EW gauge bosons $W/Z$ as $RVV$ models. $R$ may decay into $WW$, 
$ZZ$ and $ZW$ final states. This class contains 
some of the minimal extensions of the SM. We arrive at another possibility 
when one of the out going particles is $Z$ or $W$ and other particle is a BSM 
particle denoted as $RVX$ models. Remaining two possibilities arise when all 
the outgoing particles are BSM particles, denoted $RXX$ and $RXY$ models,  
where, in the later case the two final state particles are different.

\subsubsection{RVV models} The resonant particle $R$ must be a colorless 
particle as the $W$ and $Z$ bosons do not carry any color.
There are only three such possibilities for scalar $R$: singlet, 
Higgs-like $SU(2)$ doublet and a seven dimensional representation 
$(1,7,\pm2)$. The collider signature of a vector boson $R$ is similar to either 
$W'$ or $Z'$ type models.

A weak and color singlet scalar $R$ \cite{heavyhiggs} can be produced through both quark and 
gluon initiated processes. For a gluon initiated process $R$ must be 
electrically neutral. An electrically neutral scalar singlet $R$ can decay to 
$WW$ and/or 
$ZZ$ through mixing of $R$ with CP even 
Higgs ($h$) or through the following higher dimensional operators,
\begin{equation}\label{e:mod.1}
W_{\mu\nu}^a W^{\mu\nu}_a R, B_{\mu\nu} B^{\mu\nu} R,
\end{equation}
where $a=1, 2, 3$. 

% For electrically charged scalar field which $SU(3)\times 
% SU(2)$ singlets, it can talk to $W$ and  $Z$ after mixing with charged Higgs. 

For the case where $R$ is a Higgs like doublet, both the quark and gluon 
initiated production of $R$ are possible \cite{Chen2tevres, heavyhiggs, 
  WZprime}. In 
addition, the 2HDM also fits, via the charged heavy Higgs, an excess 
in the search for resonances in $W^+h$ channel\cite{Chaostealth}.

The third possibility where $R$  is an 
SU(2) seven dimensional scalar is very distinct as it contains exotic charged 
fields with charges $\pm5, \pm4,\pm3,\pm2,\pm1$ along with the neutral $R$. 

\subsubsection{RVX, RXX and RXY models:} The advantage of introducing 
non-minimal extensions of SM to explain the observed diboson excess is that 
the constraints that apply from leptonic and semi-leptonic channels can be 
partially or fully evaded with non-universal couplings with the BSM. In order 
to pass the 
experimental cuts for diboson channel, the masses of the additional final state 
particles $X$ and $Y$ 
range from $69.4{\, \rm to\,}104.8$ GeV. This also implies that they cannot be 
colored particles due to dijet constraints. As  $W, 
Z, X$ and $Y$ are all non-colored, R must also be colorless. $RXX$ models, even 
though exotic, allow an explanation for why the experiments see an excess in 
the fully hadronic final state of the diboson channel and not in the 
semi-leptonic and leptonic final 
states.

Next, considering the case where $R$ is a scalar particle which interacts with 
the quarks, there are 4 possibilities for it's EW properties,
\begin{enumerate}
 \item singlet: $\left(1,1,0\right),  \left(1,1,\pm1\right)$ , 
\item $SU(2)$ doublet like Higgs,
\item $SU(2)$ triplets $\left(1,3,\pm1\right)$, and 
\item the seven dimensional field with representation. 
$(1,7,\pm2)$.
\end{enumerate}
The final state particles $X$ and $Y$ in a subset of the models below decay 
into quark jets and mimic the massive vector boson decay-jets. $X,Y$ may 
also decay to gluons via effective couplings and may be searched for in this 
channel. These events are excluded in the diboson analyses due to cuts on 
number of charged tracks in a jet. The above 
list of representations, therefore applies to both of these particles as 
well. In the case where any of the $R,X,Y$ scalars is an SU(2) triplet, in 
order to avoid tree level corrections to the $\rho$ parameter, this scalar has 
a 0 VEV.
% These $R$ particle decays to $X$ and $Y$, and $X$ and $Y$ has to decay into 
% light quarks. 
% From Table~\ref{Tab:qRcouplings} we found the they has very 
% limited possibilities: It can be (1) a weak and color singlet, (2) a 
% SU(2) doublet and (3) a SU(2) triplet.

{\bf Singlet scalar $R$:} Consider the possibility that $R$ is a singlet 
scalar. In a model of type $RVX$, $X$ can be a singlet, 
doublet or 7 dimensional representation of SU(2). Upon symmetry breaking, 
mixing between neutral scalar bosons allows the $R\to VX$ final state. 
Similarly, through trilinear couplings arising from the scalar potential 
after electroweak symmetry breaking, $RXX$ and $RXY$ type models can also be formulated.
% 
% Construction of  $RXY$ and $RXX$ models are simple. These coupling are scalar 
% trilinear scalar coupling.  

{\bf SU(2) doublet $R$:} When $R$ belongs to an EW doublet, $RVX$ type 
models can be formulated, where the final state particle $X$ may be any 
particle from the list above. Since X has a mass similar to the $W/Z$ boson it's 
multiplet contains a light charged scalar which is constrained from the 
process $b\to s\gamma$\cite{Gambino:2001ew}. These constraints can be evaded 
with 
large quartic couplings and mass splittings within the $X$ multiplet after 
symmetry breaking. When $X$ is a singlet, a small mixing with Higgs implies 
that the coupling of $RWX$ is small and R needs a larger coupling with 
quarks/gluon to match the required production cross section. 
$RXX$ and $RXY$ type models can also be constructed with trilinear couplings 
arising from the scalar potential. When $X/Y$ has a non zero 
VEV, the mass matrix of neutral scalars has to be diagonalized with two 
eigenvalues with a large separation. This complication can be avoided with a 
zero VEV for $X$ and $Y$.
In models with multiple Higgs like doublets, the constraints due to mass 
splittings become weaker. As an example, in MSSM there are five scalar 
doublets, two from the Higgs superfields and three from lepton superfields 
which can have the required low mass scalar as well as a 
2 TeV resonant scalar\cite{Allanach:2015blv}.

{\bf SU(2) triplet $R$:} Similar to the case where $R$ is a singlet, 
$RVX,RXX$ and $RXY$ type models can be defined where the triplet particles do 
not get a VEV. In this case, masses of the zero VEV scalars are obtained as 
bare masses. 

{\bf Seven dimensional $R$:} This is an exotic case where a number of 
multi-charged particles of TeV scale  mass  arise along with the particle $R$. 
In the case of $RVX,RXX$ and $RXY$, the particle X may be a singlet, doublet 
or 7 dimensional  representation of EW 
symmetry. Additionally, $RXX$ and $RXY$ models can also be constructed with a 
triplet $X,Y$. The special case of 7 dimensional $X,Y$ is possible from the 
perspective of constraints from the $\rho$ parameter, however, a 
number of light, multi-charged scalars are present in the BSM spectrum which 
are constrained.

{\bf Gauge boson $R$:} When $R$ is a gauge boson, consider the $RXX$ coupling, 
if X is a 
gauge field, then we obtain a non minimal extension of SM with two new symmetry 
groups. This possibility is disfavored as, if $X$ belongs to a U(1) group, 
$RXX$ coupling does not exist, on the other hand, if $X$ belongs to a non 
abelian group, anomaly cancellation requires addition of multiple generations 
of new fermions.
$RVX,RXX$ and $RXY$ type models can be constructed when $X,Y$ belong to any of 
the scalar representations listed above.

\subsubsection{Some aspects of $2\to 3$ topology}
In this topology, the final state is such that two of the particles mimic the 
signal obtained from $W/Z$ bosons and the third particle is not detected with 
the CMS/ATLAS diboson analyses cuts. This requires the mass of the third hidden 
particle to be $\lesssim 100$ GeV otherwise large loss in invariant mass or 
$p_T$ asymmetry would not pass the experimental cuts.
In the case where, $R, X$ and $ Y$ are either singlets or doublets, in addition 
to the $2\to 2$ process, the $2\to 3$ process can also, in certain kinematic 
regimes, provide a signal to the diboson channel. Here we point out that all the 
models which attempt to explain the diboson excess with a 3 body final state 
suffer from a set of common constraints. With more number of massive particles 
in the final state, the phase space suppresses the cross section. The kinematic 
cuts which ensure that the two fat jets in the final state have a small relative 
$p_T$ and large separation also add further suppression in this topology. The 
couplings need to be very large to get a large enough total cross section that 
after the cuts, the required diboson excess is satisfied.

\subsection{Experimental signatures}
Dijet channel at LHC provides a lower bound on the coupling and masses of any 
BSM particle that couples to quarks or gluons. This 
constraint would apply to all the proposed $R$ particles and, via the effective 
couplings with the quarks, to the final state particle $X$ as well. Another 
interesting channel which has been discussed in detail in a previous section is 
that of associated production of $R$ with  gauge boson. This channel 
also provides crucial information about the production mechanism of $R$. In 
the case where $R$ is produced via a quark initiated process, the total charge 
of the final state leptons integrated over all the events is positive. However, 
in the case of gluon initiated process, this number would be 0. This channel 
provides an independent probe into the $RVV$ type BSM models which can 
constrain them even with low luminosity ($\sim5\, fb^{-1}$), early results from 
LHC-13.
In particular, models where the particle $X$ decays to $b\bar{b}$ or 
$\tau^+\tau^-$, the branching fraction (BF) to bottom quark should be large and 
accessible to b-tagged searches at the LHC. Also, the spin of the particle $X$ 
in the $RXX$ model can be observed through energy fractions of the sub-jets in 
reconstruction of the $W/Z$-like particles in the diboson search. Angular 
correlations of $X$ reconstructed jets can in-turn be used to determine the 
spin of $R$ once larger data points are available in the high invariant mass 
region. The BF to tau cannot be large to avoid a significant contribution to 
the semi-leptonic channel in the diboson searches where no BSM has been 
observed.
In the models where the BSM scalar transforms under a 7 dimensional 
representation of SU(2), searches for doubly charged Higgs with $T_3=0$ limit 
the mass of such a particle to $m_H^{++}\ge 322$ GeV at 95\% C.L. \cite{ATLAS:2012hi}. Due to this 
limit, a light neutral scalar $X$ cannot belong to a higher dimensional 
representation of SU(2) group since the mass splitting among the neutral and 
charged Higgs requires one neutral scalar to be heavier.

\section{Summary}
With the recently observed excess in the resonance searches in diboson channel 
at the LHC 8 TeV run, we have analyzed the phenomenological signatures of the 
BSM physics to isolate it in the observed events and provide complimentary 
signatures for LHC-13 searches. 
Analysis of the decay channels of diboson process reveals that in the early  LHC 
13 TeV run, within $\sim 5 fb^{-1}$ of data, it would be possible to confirm the 
existence of a BSM particle which fits the observed excess. In addition, we show 
that, based on the relative measurements in semi-leptonic and leptonic channels, 
as shown in Table \ref{eventcount13}, it will be possible to distinguish between 
BSM physics contributions to $WW$, $WZ$ and $ZZ$ channels.

The resonance $R$ proposed to fit the diboson excess may be generated via  a 
quark initiated or a gluon initiated process. The process of associated 
production of EW gauge bosons with $R$ can identify the initial state and 
provide an independent probe into the nature of couplings of $R$. The particle 
$R$ will also contribute to the dijet process ($p p \to R \to jj$) and the 
associated dijet process ($p p \to R V\to jjV$) . When $R$ couples with gluons, 
the final state $R+j\to jjj/VVj$ is also possible. Combining these processes 
provides a way to constrain all the couplings of $R$ with SM particles.

Since no excess has been observed in the semi-leptonic channel, we  also explore 
 a non-minimal model compatible with this observation. We add an additional 
particle denoted as $X$ with a mass $\sim 100$ GeV which primarily decay mode 
into hadronic final states. This particle can mimic the $W/Z$ signature searched 
by the experiments and satisfy the observed excess. We discuss the decay modes 
of $X$ into $\bar{q}q$, $g$g, $\bar{b}b$ and $\tau^+\tau^-$. A cut on the 
number of charged tracks ($<30$) implies that $X\to gg$ cannot be the 
primary decay mode. An analyses of the diboson search with a relaxed cut on 
number of charged tracks may add the gluon channel increasing the statistical-significance of 
the excess. A boosted b-tagging or ditau tag would also allow isolation of 
these decay channels of $X$.

In addition, we consider a different topology of models with 3 particles in the 
final state, denoted as $RXY$ models with the decay chain, $p p\to R\to V Y\to V 
V X$  where V is the $W/Z$ boson. The final state particle, $X$ can be invisible 
or can decay hadronically. The former case successfully fits the diboson excess 
only for small masses $m_X \approx 10$ GeV. For larger values of mass of 
$X$, the $p_T$ asymmetry  of the two final state gauge bosons becomes too large 
to pass the experimental cuts. This type of invisible particle can be identified 
in the boosted diboson fat jets + MET process. If $X$ dominantly decays 
hadronically, it is a part of the boosted fat jet in the final state.  Here, we 
expect to get more sub-jets in one of the fat jets, and may be identifiable by 
the two mass drops within the reconstructed jet.

In case that LHC-13 indeed finds evidence to support the presence of BSM physics in the EW sector, 
the possible models can be divided into categories based on their couplings with the qq/gg initial state and 
EW gauge bosons. We discuss the anatomy of the BSM physics which can explain the diboson excess and categorize 
the possible s-channel resonances into simplified models and list the couplings. The coupling of $R$ with the 
EW gauge bosons fixes it's SM gauge symmetries to a colorless, SU(2) singlet, doublet, or a 7 dimensional 
representation.

Finally, our results are applicable to general models which attempt to explain the diboson excess and will 
enable early detection of the BSM physics at LHC run-II. If the resonance studied here is indeed found at LHC-13, 
analyses of multiple channels which receive contributions from it will be essential tools to fix the spin, 
charge and couplings of the BSM particle(s).

\section{acknowledgements}

Work of B. Bhattacherjee is supported by Department of Science and Technology, Government of INDIA under 
the Grant Agreement numbers IFA13-PH-75 (INSPIRE Faculty Award).

\end{document}